# Varimax rotation based on gradient projection needs between 10 and more than 500 random start loading matrices for optimal performance

Anneke Cleopatra Weide & André Beauducel

Institute of Psychology, University of Bonn, Germany
September 13, 2018

**Abstract**

Gradient projection rotation (GPR) is a promising method to rotate factor or component loadings by different criteria. Since the conditions for optimal performance of GPR-Varimax are widely unknown, this simulation study investigates GPR towards the Varimax criterion in principal component analysis. The conditions of the simulation study comprise two sample sizes ($n = 100$, $n = 300$), with orthogonal simple structure population models based on four numbers of components (3, 6, 9, 12), with- and without Kaiser-normalization, and six numbers of random start loading matrices for GPR-Varimax rotation (1, 10, 50, 100, 500, 1,000). GPR-Varimax rotation always performed better when at least 10 random matrices were used for start loadings instead of the identity matrix. GPR-Varimax worked better for a small number of components, larger ($n = 300$) as compared to smaller ($n = 100$) samples, and when loadings were Kaiser-normalized before rotation. To ensure optimal (stationary) performance of GPR-Varimax in recovering orthogonal simple structure, we recommend using at least 10 iterations of start loading matrices for the rotation of up to three components and 50 iterations for up to six components. For up to nine components, rotation should be based on a sample size of at least 300 cases, Kaiser-normalization, and more than 50 different start loading matrices. For more than nine components, GPR-Varimax rotation should be based on at least 300 cases, Kaiser-normalization, and at least 500 different start loading matrices.

**Correspondence:**

Anneke C. Weide, aweide@uni-bonn.de
Institute of Psychology, University of Bonn, Kaiser-Karl-Ring 9, 53111 Bonn

# 1      Introduction

Component and factor rotation are of major relevance in behavioral research, and many rotation methods have been proposed to find interpretable solutions for loading matrices from principal component analysis (PCA) and exploratory factor analysis (Gorsuch, 1983; Harman, 1967; Mulaik, 2010). Most of these rotation methods attempt to find simple structure. Bernaards and Jennrich (2005) propose a generalized algorithm for analytic factor rotation that can perform probably most available rotation methods towards simple structure by means of gradient projection (GPR). The approach is very promising, because it unifies different rotation criteria under a single algorithm. Making it even more appealing, GPR can be performed in both commercial and non-commercial statistical software. GPR implementations of most rotation criteria are available for SPSS Matrix, SAS PROC IML, Matlab, Splus, and R (http://www.stat.ucla.edu/research/gpa). To this point, the mathematical background of the algorithm and its refinements have been comprehensively and extensively described (Bernaards & Jennrich, 2005; Jennrich, 2001, 2002). The authors also provide demonstrations on how to use GPR with exemplary loading matrices. However, simulation studies need to be conducted to evaluate the performance of GPR for applications of common usage. We therefore provide a simulation study on GPR by investigating its performance for different conditions of rotation towards orthogonal simple structure. Orthogonal simple structure can effectively be recovered by orthogonal rotation criteria, amongst which the Varimax criterion (Kaiser, 1958) is probably the most popular and accepted one (Browne, 2001; Fabrigar, Wegener, MacCallum, & Strahan, 1999). We decided to consider the Varimax criterion in the present study, because it is still one of the most common criteria for orthogonal rotation. For example, searching for "Varimax" on GoogleScholar yielded 196,000 hits, as compared to 5,670 for "Quartimax", 3,490 for "Equamax", and 172 entries for "Parsimax". Even though perfect orthogonality is rather unlikely, Varimax also exceeds the popularity of oblique rotation criteria, such as Promax (40,100 hits), Oblimin (37,000 hits), or Quartimin (1,630 hits). As one may assume that the relative popularity of Varimax rotation is due to the lack of a backward limit for the time frame in this search, we performed an additional GoogleScholar search for publications in the time frame between 2014 and 2017. We got 23,400 hits when entering "Varimax" into a search from 2014 to 2017. When "Geomin", a more recent method for oblique and orthogonal rotation (Browne, 2001), was entered for the same time frame, we got 1,660 hits. Thus, Varimax rotation is still one of the most used rotation methods, even when many other alternative rotation methods are meanwhile available (Browne, 2001). However, scientists that use Varimax seem to rely on procedures that can be performed by pushing a button in commercial statistical software like SPSS or SAS. Entering the combination of "Varimax" and "SPSS" yielded 17,400 hits, and we got 4,930 hits for the combination of "Varimax" and "SAS" in GoogleScholar. Meanwhile, the combination of "Varimax" and "gradient projection" resulted in only 41 hits in a GoogleScholar search in the abovementioned time frame (all searches were performed on September 13, 2018, with the quotation mark operator). Hence, GPR has not yet reached the popularity that could be expected given its promising simplicity and generalizability, and it faces the challenge of competing against readily available procedures in established software. If GPR should be used by more scientists, it is important that it can be supported by simulations that investigate its performance for the most relevant rotation criteria, also in comparison to procedures that are implemented in commercial software. Since SPSS is one of the most popular software tools for factor rotation and the Varimax criterion is one of the most popular rotation criteria, this study is devoted to investigating the implementation of GPR in SPSS for the Varimax criterion and comparing it to the built-in SPSS procedure. As for the choice between factor analysis and PCA, we investigate GPR-Varimax performance for the rotation of components because the component model is simpler and does not require the estimation of error factors (Harman, 1967). Thereby, the present simulation does not depend on the precision of different methods for the estimation of factor loadings and communalities and focuses on the precision of the rotation method only.

In order to understand the approach of the present simulation study, we give a brief description of the GPR algorithm in factor and component rotation (Bernaards & Jennrich, 2005; Jennrich, 2001, 2002). More detailed descriptions of the algorithm can be found elsewhere (Mulaik, 2010). The idea of GPR is that any rotation towards simple structure relies on an optimization (minimization or maximization) criterion, where constraints are placed on some parts of the optimization function. The gradient projection algorithm can be used to solve such constraint optimization problems. A rotation of an initial $m \times k$ loading matrix **A** (e.g. unrotated loadings) is given by

$$\mathbf{\Lambda} = \mathbf{AT} \tag{1}$$

where **T** is a $k \times k$ transformation matrix with columns of unit length (i.e. its column sums of squares add up to 1). Let **Φ** be the correlation matrix between rotated factors or components. For an orthogonal rotation, the rotated components are required to remain uncorrelated, such that the $k(k-1)$ non-diagonal elements of **Φ** are constrained to be zero, resulting in

$$\mathbf{\Phi} = \mathbf{TT'} = \mathbf{I}. \tag{2}$$

The optimization criterion $Q$ in factor and component rotation is a function of the rotated loading matrix **Λ** and thereby a function of **T**, denoted by

$$f(\mathbf{T}) = Q(\mathbf{\Lambda}). \tag{3}$$

For example, the Varimax criterion seeks to maximize the variance of squared loadings (Kaiser, 1958) and is therefore a function of the transformation matrix **T**. The algorithm searches for a minimum of $f(\mathbf{T})$, such that the Varimax criterion would be the negative of $Q(\mathbf{\Lambda})$. The algorithm uses the negative gradient **G** of $f(\mathbf{T}) = Q(\mathbf{\Lambda})$, given by

$$\mathbf{G} = -\left(\mathbf{\Lambda}' \frac{\partial Q}{\partial \mathbf{\Lambda}} \mathbf{T}^{-1}\right)'. \tag{4}$$

Each rotation criterion has its own expression for $\partial Q$. The algorithm starts with an initial **T** from the manifold **T** that comply with the constraints. Then, **T** is moved by its negative gradient with step length $\alpha$ to find a matrix **M** by

$$\mathbf{M} = \mathbf{T} - \alpha \mathbf{G}. \tag{5}$$

Next, **M** is projected back onto **T** by normalizing its columns to unit length. The projection is denoted by ~**T**. After each step, the rotation criterion $f(\text{\textasciitilde}\mathbf{T})$ is evaluated and compared to the previous $f(\mathbf{T})$. For a sufficiently small $\alpha$, the algorithm is strictly descending, and

$$f(\text{\textasciitilde}\mathbf{T}) < f(\mathbf{T}). \tag{6}$$

Then, **T** is replaced with ~**T**, and the algorithm starts again. If $f(\text{\textasciitilde}\mathbf{T}) \geq f(\mathbf{T})$, $\alpha$ needs to be reduced (e.g. halved), until $f(\text{\textasciitilde}\mathbf{T}) < f(\mathbf{T})$. The algorithm continues until it converges, that is, when $f(\mathbf{T})$ becomes minimal.

For the initial loading matrix $\mathbf{\Lambda} = \mathbf{AT}$, we can insert any start transformation matrix $\mathbf{T}$ that complies with the constraints. For example, we can use the $k \times k$ identity matrix (Jennrich, 2001) or a random matrix whose columns have unit length (Bernaards & Jennrich, 2005; Mulaik, 2010) For the latter, Bernaards and Jennrich (2005) recommend running the GPR algorithm several times with multiple random start matrices to identify local optima. Local optima are of major relevance in analytic rotation, because the algorithms search a minimum on curvilinear, complex loss-functions with many ups and downs (Rozeboom, 1992). Hence, beginning with a particular start transformation matrix $\mathbf{T}$ does not guarantee to find the global optimum for the rotation criterion. Addressing this issue, multiple random starts are often used in demonstrations of analytic rotation algorithms. For example, Kiers (1994) used 20 random start matrices to present the Simplimax method, just like Browne (2001) in his comparative overview on different analytic rotation methods. Recently, Hattori, Zhang, and Preacher (2017) published a paper on Geomin rotation, in which they identified local optima by iterating across 100 random start matrices. The code for multiple random start matrices for GPR is available online (http://www.stat.ucla.edu/research/gpa). However, when the algorithm is performed on only a single start loading matrix, the success of GPR in revealing a global optimum heavily depends on $\mathbf{A}$ because it is inserted without any alterations ($\mathbf{\Lambda} = \mathbf{AT} = \mathbf{AI} = \mathbf{A}$). If $\mathbf{A}$ is an unrotated loading matrix, we can assume that the local optimum found by the GPR implementation is unlikely to be the global optimum of the rotation criterion. This could explain why Dien (2010) found less favorable results for Geomin and Oblimin rotation of components when he applied the GPR implementations to simulated EEG data (event-related potentials) and compared them to other rotation methods.

Even though the existence of local optima in GPR has been acknowledged, the information on the number of random starts necessary to find a global optimum for any of the rotation criteria is scarce. This simulation study therefore provides information on the number of random starts necessary to find a global maximum for Varimax rotation based on PCA components in the case of perfect orthogonal simple structure in the population. We define that a global optimum is reached when rotation performance does not improve substantially with further iterations of the random start loading matrices. Rotation performance thereby refers to the Varimax criterion and congruence of rotated components with population components. Hence, rotation performance of GPR-Varimax becomes optimal when both the Varimax criterion and population congruence reach stationarity. Apart from stationarity, we include a second reference for rotation performance of GPR-Varimax by comparing results for different numbers of random start loading matrices to the built-in Varimax-rotation of the IBM SPSS software package. We want to investigate GPR-Varimax performance for different conditions and will therefore vary sample size and the number of components to be rotated. For the manipulated conditions, we will assess how many iterations of initial loading matrices are necessary until GPR-Varimax reaches optimal (stationary) performance. Hence, we examine the performance of GPR-Varimax in PCA under different conditions of orthogonal simple structure. Based on the results of our simulation study, we provide recommendations on how many iterations of random start matrices are necessary to reach optimal (stationary) rotation performance under these conditions.

## 2   Material and Methods

We used IBM SPSS Version 25 for all analyses. GPR-Varimax was performed with the SPSS matrix implementations of the GPR algorithm (Bernaards & Jennrich, 2005). We also performed the built-in Varimax-rotation of SPSS, which is available with the SPSS command *FACTOR*, on the data (SPSS-Varimax, based on 250 iterations). The conditions that we manipulated refer to number of true components in the population ($k = 3, 6, 9$, and $12$) and the number of variables respectively ($m = 18$, 36, 54, and 72), sample size ($n = 100$ and $n = 300$), and whether Varimax-rotation was based on Kaiser-normalized loadings or not.

## 2.1 Population component models

We used population components instead of population factors because they allow for a fairer evaluation of GPR-Varimax rotation of components in the samples. To obtain population components, we generated population data sets on the basis of orthogonal common and error factors and submitted these population data sets to PCA. We used $k/m = 6$, so that there were six variables with main loadings on each factor. All populations were modeled with perfect simple structure. Thus, each common factor had main loadings of $a = .50$ on six variables and zero loadings on all other variables. Loadings for error factors were $d = \sqrt{1 - .50^2}$. We followed a procedure described by Grice (2001) to generate data sets for finite populations based on common and error factors. Thereby, we first used the SPSS Mersenne Twister random number generator to generate normally distributed, z-standardized random data $\mathbf{X}$ containing $k + m$ preliminary factor scores. We then performed PCA on these data, again extracting $k + m$ components without rotating them. Since component extraction in PCA is based on orthogonality, this step ensures that the final factors, common and error, are all based on perfectly uncorrelated random variables to fulfil the requirements of population data. We saved the $k + m$ component scores, which served as random variables for the $k$ common factors $\mathbf{F}$ and $m$ error factor scores $\mathbf{U}$. We then used an SPSS Matrix script in order to generate a finite population of observed variables by the common factor model $\mathbf{Z} = \mathbf{FA'} + \mathbf{UD'}$ (Gorsuch, 1983; Grice, 2001). The population data sets consisted of 100,000 cases for all populations that comprised 1,000 samples with $n = 100$ and 300,000 cases for all populations of 1,000 samples with $n = 300$. Finally, population component loadings were computed by performing a PCA with the built-in Varimax rotation in SPSS on the population data. Population component loadings were $a^* = .61$. We checked this result by re-computing the corresponding factor loadings by $a = \sqrt{a^*(1 - (1 - h^2)/\lambda^*}$ , where $h^2$ denotes the communality of the variables (that were all equal in the population), and $\lambda^*$ denotes the eigenvalue of the component (Harris, 2001, p. 417). We obtained the original factor loadings of $a = .50$ for corresponding components of $a^* = .61$.

## 2.2 Analysis of simulated sample data

From the population data, 1,000 samples were drawn with either $n = 100$ or $n = 300$. Data for the samples were submitted to correlation-based PCA in SPSS. We extracted a fixed number of components in the samples corresponding to the correct number of population components. Components of each sample were rotated with GPR-Varimax and SPSS-Varimax, both performed with and without Kaiser-normalization (Kaiser, 1958). For initial loading matrices $\mathbf{\Lambda} = \mathbf{AT}$ in GPR-Varimax, we used the $k \times k$ identity matrix and random matrices as start transformation matrices $\mathbf{T}$. For the computation of random start loading matrices for GPR-Varimax rotation, we used the SPSS Mersenne Twister random number generator in order to generate random transformation matrices. The unrotated component loading matrices $\mathbf{A}$ were post-multiplied by the random transformation matrices in order to generate random loading matrices. We allowed for six numbers of iterations to obtain random start loading matrices ($q = 1, 10, 50, 100, 500, 1,000$) in order to find a maximum for the Varimax criterion.

We assessed rotation performance by computing the Varimax criterion (Kaiser, 1958) $v$ for all rotations and by comparing GPR-Varimax loadings to population component loadings. Proximity of the Varimax-rotated sample loadings to the respective population components was computed by congruence coefficients $c$ for rotated sample loadings and corresponding population component loadings (Korth & Tucker, 1975; Tucker, 1951). For each sample, congruence coefficients were averaged across components. Moreover, deviation of the sample loadings from population component loadings was calculated by the root mean square error (RMSE) based on the squared differences between rotated and population loadings (see Supplement). For each condition and method (GPR-Varimax, SPSS-Varimax, with and without Kaiser-normalization), we averaged values for the three criteria ($v$, $c$, RMSE) of rotation performance across all 1,000 samples.

## 3   Results

We focused on two aspects in our analyses: First, we examined effects of the conditions (number of factors $k$, sample size $n$, and Kaiser-normalization) and start loading matrices (unrotated loadings with identity transformation matrix or random loadings with random transformation matrices) on rotation performance of GPR-Varimax in general. Second, we investigated for each condition, how many iterations of random start loading matrices were necessary to reach a stationary point and thereby optimal results for the Varimax criterion and for congruence with population component loadings.

### 3.1   Cut-offs for stationarity and equality of results

For the investigation of stationarity, we defined cut-offs for sufficient equality between two solutions of consecutive number of iterations (e.g. $q = 10$ vs. $q = 50$). We applied the same criterion to evaluate equality of GPR-Varimax and SPSS-Varimax results as for stationarity. For all comparisons, we decided on a conjunctive condition for congruence with population loadings and the Varimax criterion. Cut-offs were chosen to match the scaling of the values. Mean congruence coefficients in the samples ranged from $c = .627$ to $c = .984$, where most differences became apparent at the second and third decimal. The mean Varimax criterion in the samples ranged from $v = .0083$ to $v = .0311$, and differences were mostly found at the third and fourth decimal. Therefore, we considered results to be equal when both the difference between mean congruence coefficients was $\leq .001$ and the difference between the mean Varimax criterion was $\leq .0001$. In other words, results were considered equal when congruences differed by no more than .001 and the Varimax criterion differed by no more than .0001. For stationarity, when results between two consecutive numbers of iterations were found to be equal, we considered the smaller number of iterations to suffice for optimal (stationary) results. Table 1 uses these cut-offs to give an overview on how many iterations of start loading matrices were necessary for GPR-Varimax rotation to reach stationarity. Table 2 refers to the comparison of GPR-Varimax with the built-in SPSS procedure for Varimax rotation. It gives the minimum number of start loading matrices that were necessary for GPR-Varimax to reach SPSS performance according to the cut-offs.

### 3.2   Effects of condition

With respect to condition effects, it took more iterations of start loadings for GPR-Varimax to reach a stationary point when more components were rotated (see Figures 1 to 8). Similarly, GPR-Varimax performed worse in smaller samples as compared to larger samples, and GPR-Varimax needed more iterations of start loadings in small samples than in large samples. Using Kaiser-normalized loadings for rotation enhanced rotation performance of GPR-Varimax considerably. For Kaiser-normalized loadings, fewer iterations of random start loading matrices were necessary than for non-normalized loadings (see Table 1). This effect was especially prominent and relevant for larger numbers of population factors and rotated components and for smaller samples. For all conditions, it was better to use at least 10 different random start loading matrices as compared to using unrotated component loadings (= identity transformation matrix) as start loading matrices (see Figures 1 to 8). For most conditions, it was even slightly advantageous to use only a single random transformation matrix rather than the identity matrix. Thus, even slight changes to unrotated loadings are favorable for GPR-Varimax. However, both the identity transformation matrix and single random transformation matrices resulted in local optima for the Varimax criterion. Consequently, the resulting rotated loadings did not optimally reproduce population simple structure.

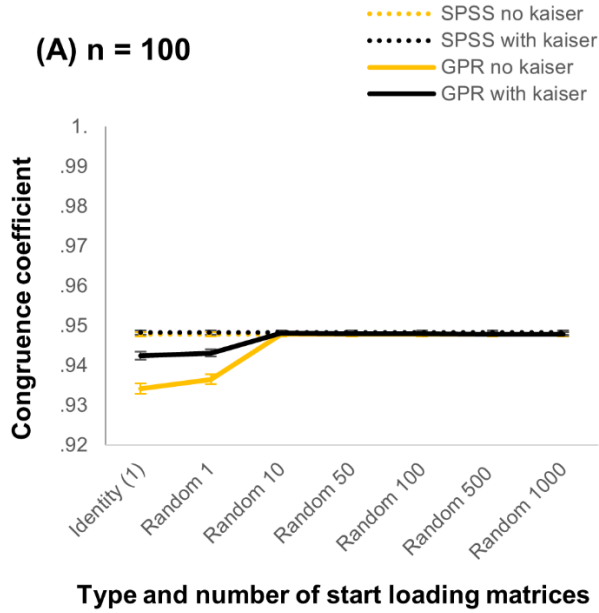
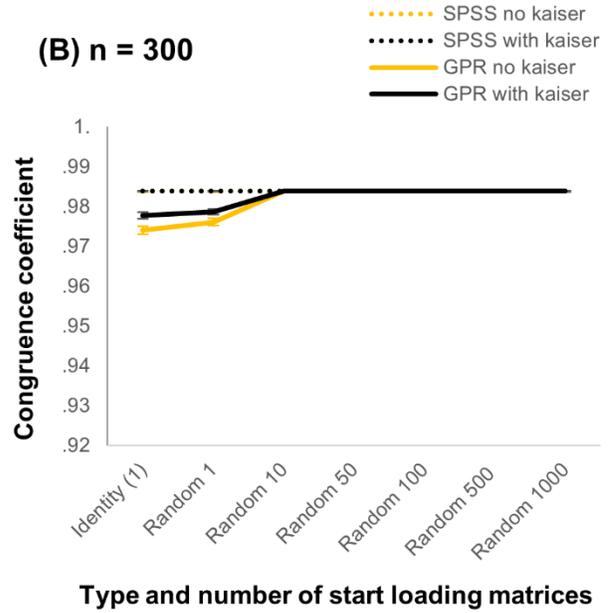

*Figure 1.* Mean congruence (and standard errors) with population components for the rotation of 3 components. *Type of start loadings* refers to the start transformation matrix **T** by which the unrotated loadings **A** are post-multiplied before GPR-Varimax rotation. *Identity* means that **A** was inserted without alterations. The minimum value of the axis is the minimal empirical value for mean congruence from the simulation - .01.

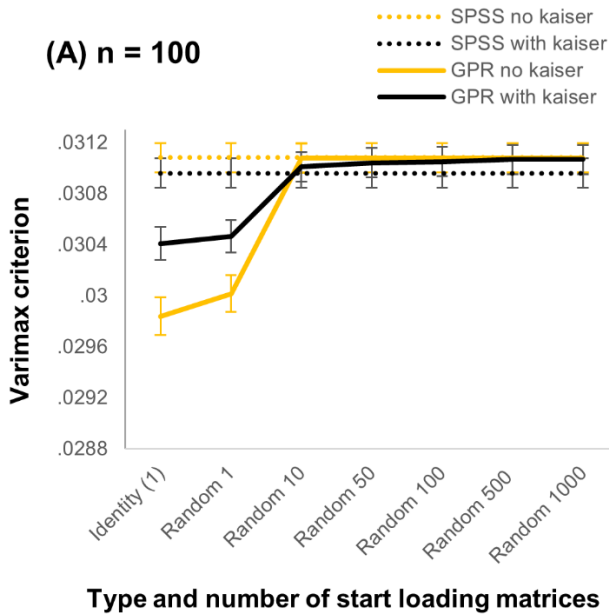
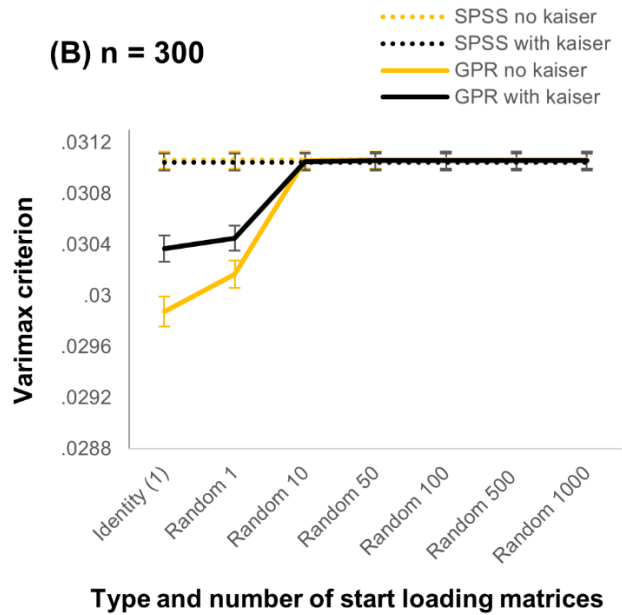

*Figure 2.* Mean Varimax criterion (and standard errors) for the rotation of 3 components. *Type of start loadings* refers to the start transformation matrix **T** by which the unrotated loadings **A** are post-multiplied before GPR-Varimax rotation. *Identity* means that **A** was inserted without alterations. The maximum value of the axis is the Varimax criterion computed for the corresponding population components. The minimum value of the axis is the minimal empirical value for the mean Varimax criterion from the simulation - .001.

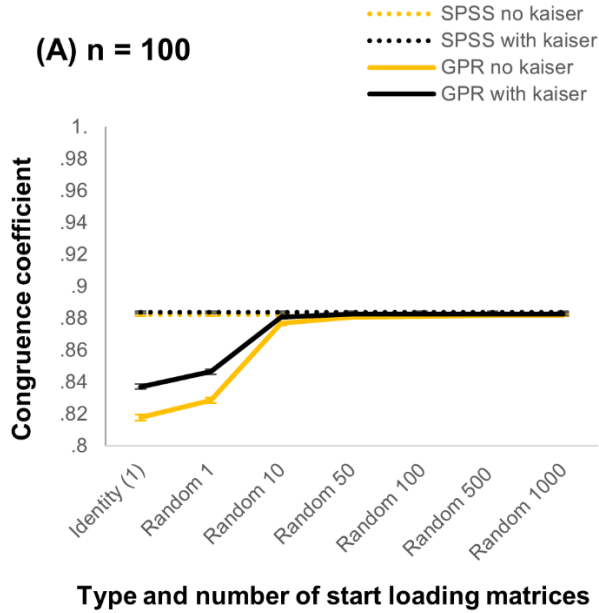 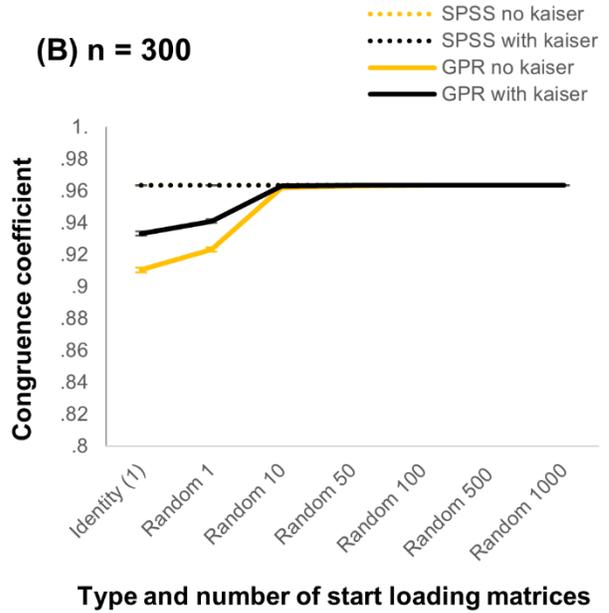

*Figure 3.* Mean congruence (and standard errors) with population components for the rotation of 6 components. *Type of start loadings* refers to the start transformation matrix **T** by which the unrotated loadings **A** are post-multiplied before GPR-Varimax rotation. *Identity* means that **A** was inserted without alterations. The minimum value of the axis is the minimal empirical value for mean congruence from the simulation - .01.

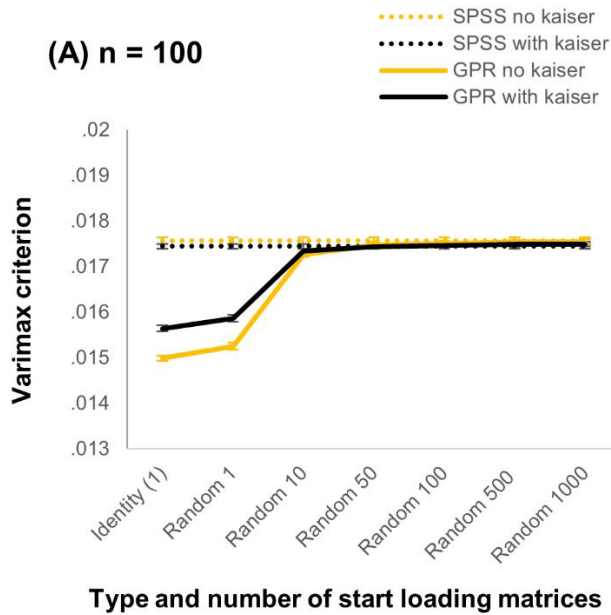 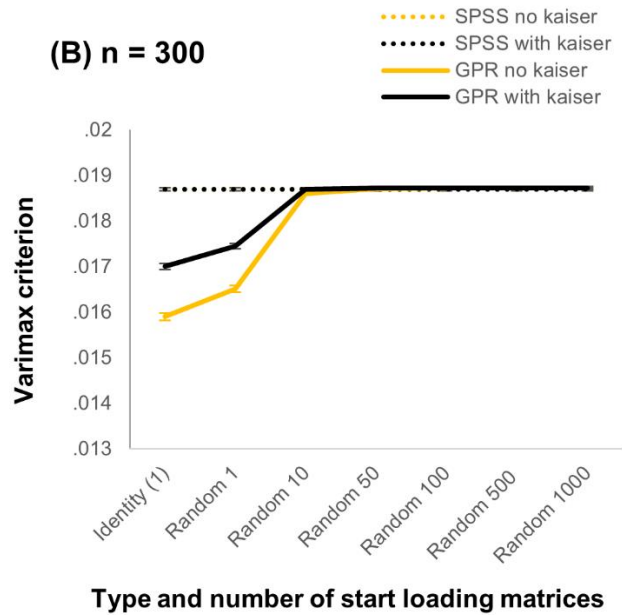

*Figure 4.* Mean Varimax criterion (and standard errors) for the rotation of 6 components. *Type of start loadings* refers to the start transformation matrix **T** by which the unrotated loadings **A** are post-multiplied before GPR-Varimax rotation. *Identity* means that **A** was inserted without alterations. The maximum value of the axis is the Varimax criterion computed for the corresponding population components. The minimum value of the axis is the minimal empirical value for the mean Varimax criterion from the simulation - .001.

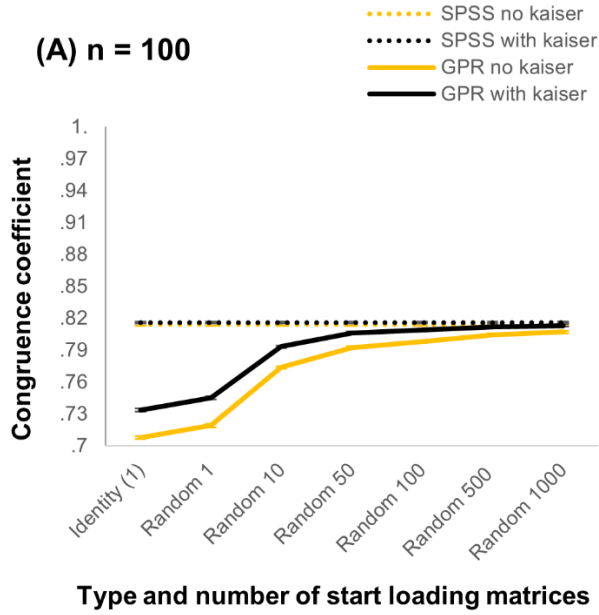 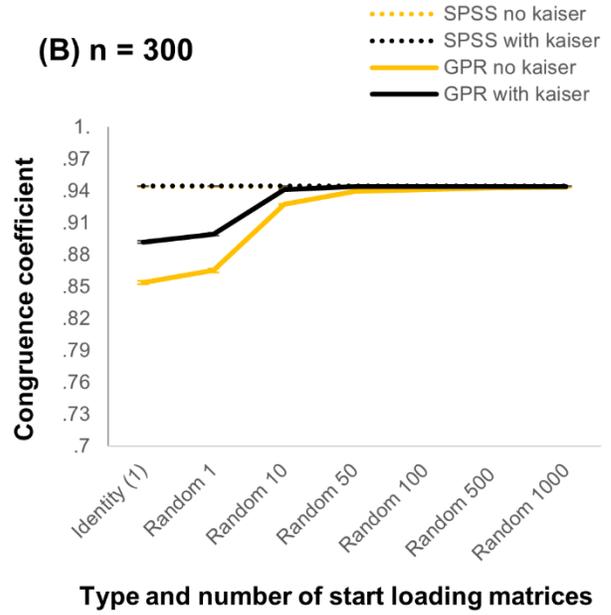

*Figure 5.* Mean congruence (and standard errors) with population components for the rotation of 9 components. *Type of start loadings* refers to the start transformation matrix **T** by which the unrotated loadings **A** are post-multiplied before GPR-Varimax rotation. *Identity* means that **A** was inserted without alterations. The minimum value of the axis is the minimal empirical value for mean congruence from the simulation - .01.

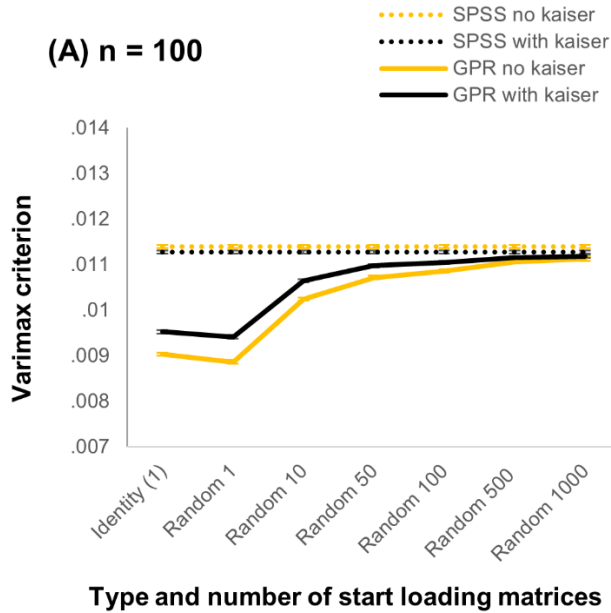 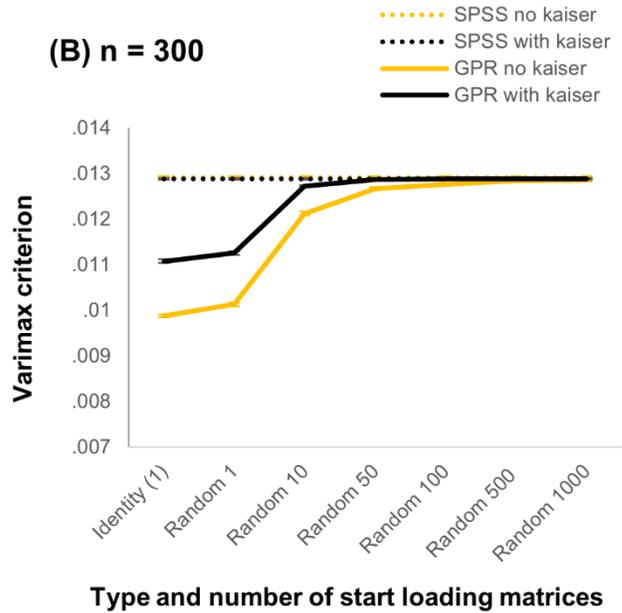

*Figure 6.* Mean Varimax criterion (and standard errors) for the rotation of 9 components. *Type of start loadings* refers to the start transformation matrix **T** by which the unrotated loadings **A** are post-multiplied before GPR-Varimax rotation. *Identity* means that **A** was inserted without alterations. The maximum value of the axis is the Varimax criterion computed for the corresponding population components. The minimum value of the axis is the minimal empirical value for the mean Varimax criterion from the simulation - .001.

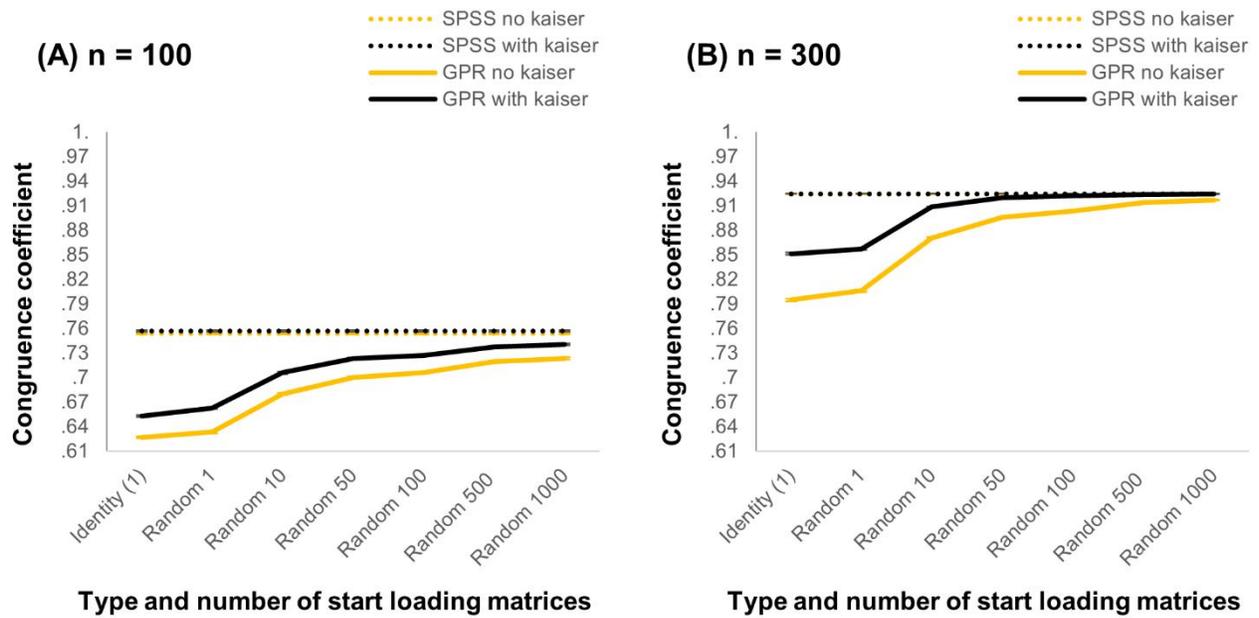

*Figure 7.* Mean congruence (and standard errors) with population components for the rotation of 12 components. *Type of start loadings* refers to the start transformation matrix **T** by which the unrotated loadings **A** are post-multiplied before GPR-Varimax rotation. *Identity* means that **A** was inserted without alterations. The minimum value of the axis is the minimal empirical value for mean congruence from the simulation - .01.

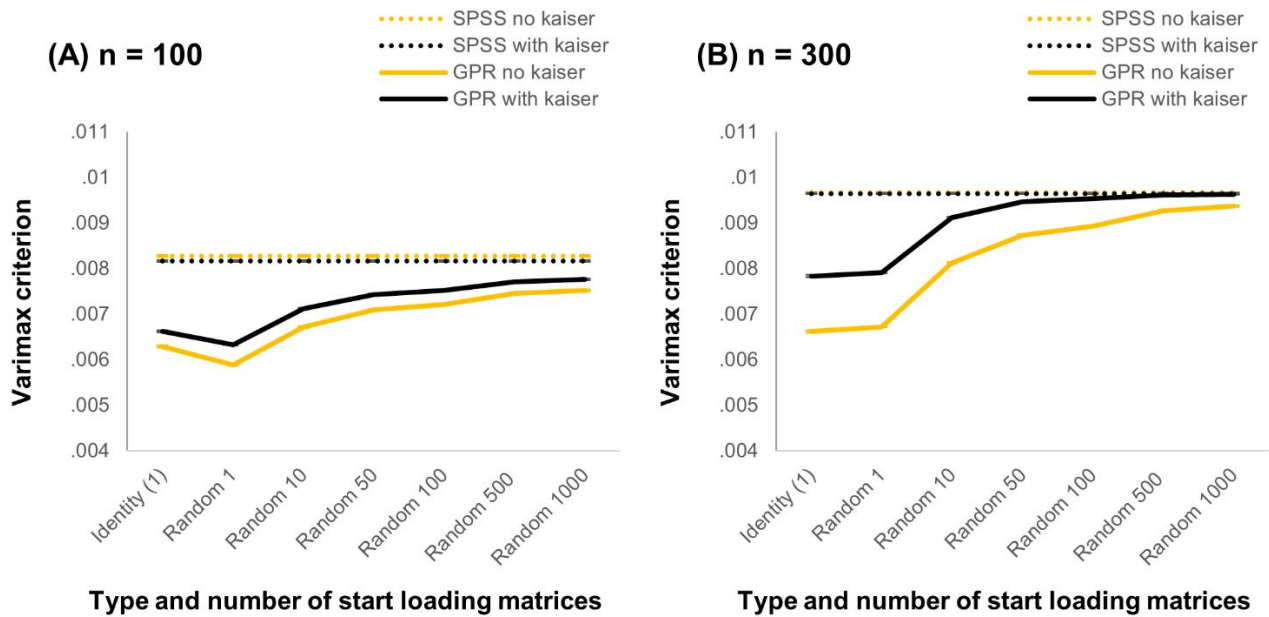

*Figure 8.* Mean Varimax criterion (and standard errors) for the rotation of 12 components. *Type of start loadings* refers to the start transformation matrix **T** by which the unrotated loadings **A** are post-multiplied before GPR-Varimax rotation. *Identity* means that **A** was inserted without alterations. The maximum value of the axis is the Varimax criterion computed for the corresponding population components. The minimum value of the axis is the minimal empirical value for the mean Varimax criterion from the simulation - .001.

### 3.3 Minimum number of start loadings for optimal performance of GPR-Varimax

For each condition, we identified the number of start loading matrices that was necessary for GPR-Varimax rotation to reach optimal performance (see Table 1). For three population factors, it was sufficient to use 10 iterations of random start loading matrices to reach optimal performance in all sub-conditions (see Figure 1 and 2). For six population factors, sample size and Kaiser-normalization showed slight effects (see Figure 3 and 4). However, when the cut-offs were used in the $k = 6$ condition, Kaiser-normalization showed an effect only for $n = 300$ (see Table 1). For $k = 6$ and $n = 100$, the start loading matrix had to be iterated 50 times in order to reach optimal performance for both Kaiser-normalized and non-normalized loadings. GPR-Varimax also needed 50 iterations in the $k = 6$, $n = 300$ condition when non-normalized loadings were rotated, whereas 10 iterations sufficed for Kaiser-normalized loadings. For the rotation of nine components, effects of sample size and Kaiser-normalization increased (see Figure 5 and 6). Hence, the minimum number of iterations varied between 50 for Kaiser-normalized loadings in large samples ($n = 300$) and more than 500 for loadings in small samples ($n = 100$). In other words, for $k = 9$ and $n = 100$, GPR-Varimax did not meet the criteria for stationarity within 500 iterations of random start matrices for neither Kaiser-normalized nor non-normalized loadings. The results were most pronounced for the rotation of 12 components (see Figure 7 and 8), where only in the sub-condition with $n = 300$ and Kaiser-normalized loadings GPR-Varimax reached stationarity within a minimum of 500 iterations. For all other sub-conditions with $k = 12$, differences between 500 and 1000 iterations were too big, such that 500 iterations were not yet enough to find a global optimum.

*Table 1*

*Minimum number of random start values in GPR-Varimax necessary to reach stationarity in rotation performance*

| | N = 100 | | N = 300 | |
|---|---|---|---|---|
| Number of components | No Kaiser-normalization | With Kaiser-normalization | No Kaiser-normalization | With Kaiser-normalization |
| 3 | 10 | 10 | 10 | 10 |
| 6 | 50 | 50 | 50 | 10 |
| 9 | > 500 | > 500 | 500 | 50 |
| 12 | > 500 | > 500 | > 500 | 500 |

*Note.* GPR = gradient-projection based rotation. Rotation after principal component analysis of simulated data from populations with perfect simple structure of orthogonal common components. Number of iterations of random start values in GPR-Varimax were 1, 10, 50, 100, 500, and 1,000. Results of two consecutive numbers of iterations (e.g. 50 vs. 10) were considered equal when the difference between mean congruence with population components was ≤ .001 and the difference between the mean Varimax criterion was ≤ .0001. Both conditions had to be fulfilled. Since two consecutive numbers of iterations were necessary to assess stationarity, highest results refer to the 1,000 – 500 difference.

### 3.4 Comparison of GPR-Varimax with SPSS-Varimax

Supporting the findings for optimal (stationary) performance, plateaus of stationarity for GPR-Varimax were close to the level of results for Varimax-rotated loadings obtained by the SPSS procedure with the command *FACTOR* (see Figures 1 to 8). As can be seen in the figures, SPSS-Varimax rotations resulted in equal or higher mean congruencies of the sample components with the corresponding population components. Hence, SPSS-Varimax rotation may be considered as a benchmark for the evaluation GPR-Varimax rotation. Accordingly, the minimum number of iterations in GPR-Varimax necessary to reach SPSS-Varimax performance (Table 2) shows a similar pattern as the minimum number of iterations that is necessary in order to reach stationarity (Table 1). For $k = 3$, GPR-Varimax reached SPSS-Varimax performance within 10 iterations of start loading matrices, just like for stationarity. For $k = 6$, results were the same as for stationarity in the $n = 300$ sub-condition, but more iterations were required than for stationarity in the $n = 100$ sub-condition. Where stationarity was reached within 50 iterations of start loadings for $k = 6$ and $n = 100$, GPR-Varimax required 100 iterations for non-normalized and 500 iterations for Kaiser-normalized loadings to reach SPSS performance. For larger numbers of components, the comparison for up to 1,000 iterations in GPR-Varimax became relevant. For $k = 9$ and $n = 100$, iterating 1,000 times across start loadings was not yet enough to reach SPSS results. Like for stationarity, 50 iterations of start loadings in GPR-Varimax were enough to reach SPSS performance for $k = 9$ and $n = 300$ when Kaiser-normalized loadings were rotated. In contrast, when loadings were non-normalized, 1,000 iterations of start loading matrices were necessary for GPR-Varimax to reach SPSS performance. The SPSS-GPR comparison paralleled the results for stationarity for the rotation of 12 components. GPR-Varimax reached SPSS performance within 500 iterations only for the sub-condition of $n = 300$ and Kaiser-normalization when 12 components were rotated. For the other sub-conditions with 12 components, GPR-Varimax did not reach SPSS-Varimax performance even when GPR start loadings were iterated 1,000 times.

*Table 2*

*Minimum number of random start values in GPR-Varimax necessary to reach SPSS-Varimax performance*

| Number of components | N = 100 | | N = 300 | |
| --- | --- | --- | --- | --- |
| | No Kaiser-normalization | With Kaiser-normalization | No Kaiser-normalization | With Kaiser-normalization |
| 3 | 10 | 10 | 10 | 10 |
| 6 | 100 | 500 | 50 | 10 |
| 9 | > 1000 | > 1000 | 1000 | 50 |
| 12 | > 1000 | > 1000 | > 1000 | 500 |

*Note.* GPR = gradient-projection based rotation. Rotation after principal component analysis of simulated data from populations with perfect simple structure of orthogonal common components. Number of iterations of random start values in GPR-Varimax were 1, 10, 50, 100, 500, and 1,000. Results of GPR-Varimax were considered equal to SPSS-Varimax when the difference between mean congruence with population components was $\leq$ .001 and the difference between the mean Varimax criterion was $\leq$ .0001. Both conditions had to be fulfilled.

Effects for deviation from population loadings, computed by the RMSE statistic, mirrored the effects for proximity to population loadings (congruence coefficients) and for the Varimax criterion. They can be inspected in the Supplement of this paper.

## 4    Discussion

Building on the discussion of multiple local optima in analytic factor and component rotation (Bernaards & Jennrich, 2005; Mulaik, 2010; Rozeboom, 1992), we examined the performance of GPR-Varimax in PCA to ascertain how many iterations of initial loading matrices are necessary for optimal performance in different conditions. Thus, Varimax-rotation, the probably still most popular method for orthogonal rotation towards orthogonal simple structure (Browne, 2001; Fabrigar et al., 1999), was investigated for population models that were based on perfect orthogonal simple structure. Optimal performance was defined as a point where no substantial improvements of the Varimax-criterion and of the congruence with population simple structure was possible by means GPR-Varimax rotation of further random start loading matrices. We also compared GPR-Varimax performance with the performance of the built-in Varimax-rotations in SPSS that are available with the command *FACTOR*.

### 4.1    Summary and interpretation of results

We found that rotation performance was always better when multiple random start matrices were used instead of an identity start transformation matrix. Depending on condition, and mostly on the number of rotated components, between 10 and more than 500 iterations of start loading matrices in GPR-Varimax were necessary to reach optimal performance. Rotation performance improved when loadings were Kaiser-normalized before rotation, especially when the number of components was large. We found that 10 iterations of start loading matrices were enough for the rotation of three components for both sample sizes ($n = 100$ and $n = 300$) and independent of Kaiser-normalization. For six components, 10 iterations sufficed only for $n = 300$ and Kaiser-normalization, whereas 50 iterations were required for the other sub-conditions to reach optimal results. The effects of Kaiser-normalization became more prominent for the rotation of nine and 12 components. The largest variation in the minimum number of iterations was found for the rotation of nine components. It ranged between 50 for $n = 300$ and Kaiser-normalized loadings and more than 500 for $n = 100$. When 12 components were rotated, it was possible to find optimal results within 500 iterations of start matrices only in the sub-condition with $n = 300$ and Kaiser-normalization. Overall, SPSS-Varimax rotations of sample components resulted in equal or larger congruences with the corresponding population components, so that they may be considered as a benchmark for the evaluation of GPR-Varimax rotation. The comparison between GPR-Varimax and SPSS-Varimax paralleled the results for stationarity. Hence, the minimum number of iterations in GPR-Varimax for stationarity was similar to the minimum number of iterations necessary to reach SPSS-Varimax performance. Also,
whenever stationarity could not be reached within 500 iterations of start loadings, GPR-Varimax did not reach SPSS-Varimax performance within 1,000 iterations. This is particularly relevant for the rotation of nine or more components and small samples.

The interpretation of our findings is limited to the choices we made regarding the number of iterations of random start loading matrices, the number of components to be rotated, and sample size. We decided that start loading matrices should be iterated 1, 10, 50, 100, 500, and 1,000 times to evaluate at what point GPR-Varimax reached stationarity and optimal results. This exceeds the number of iterations allowed in previous demonstrations of factor rotation where multiple local optima were of concern (Browne, 2001; Hattori et al., 2017; Kiers, 1994; Trendafilov & Jolliffe, 2006). While 10 to 50 iterations seem enough for the rotation of three or six components, 500 iterations did not yet suffice for optimal rotation performance of GPR-Varimax in some sub-conditions with nine and 12 components. The number of components for which we investigated performance of GPR-Varimax covers the numbers of facets embedded in the most popular theories on personality. The main discussion revolves around whether personality consists of three (Eysenck, 1991; Eysenck, Barrett, Wilson, & Jackson, 1992), five (Costa & McCrae, 2006) or six (Lee & Ashton, 2004) independent components or factors. For the rotation of more than 12 components or factors, as would be necessary when instruments like the 16-Personality-Factor-Inventory (Cattell, 1956) are used, we assume that effects would become more pronounced. With regards to sampling error, we used $n = 100$ as the minimum sample size recommended for factor analysis in the literature (Gorsuch, 1983). The second sample size of $n = 300$ exceeds most recommendations for minimum sample sizes and has been classified as *good* (MacCallum, Widaman, Zhang, & Hong, 1999). However, since for larger samples, rotation performance was better, and the number of necessary iterations was smaller, it seems advantageous to perform GPR-Varimax in samples of $n = 300$ or larger. It is likely that GPR-Varimax reaches optimal results within fewer number of iterations than presented here when it is performed on data from larger samples. Sample size also affected the height of the plateaus for congruences and the Varimax criterion, which we found in our analyses of stationarity (see Figures 1 to 8). With respect to congruences, it has been suggested that values between $c = .85$ and $c = .94$ indicate fair similarity between components or factors, and components with a congruence of $c = .95$ can be regarded as equal (Lorenzo-Seva & ten Berge, J. M. F., 2006). As can be seen in Figures 1, 3, 4, and 6, most plateaus of stationarity fall into this range. Hence, when GPR-Varimax reaches stationarity, we can assume that it recovers orthogonal population components to an acceptable degree. Only when GPR-Varimax was performed in small samples ($n = 100$) on nine and 12 components, the plateau for optimal congruence was below this threshold. However, also the built-in Varimax rotation in SPSS did not exceed the threshold for congruences in these sub-conditions.

## 4.2 Recommendations on using GPR-Varimax

The results of the simulations are a basis for recommendations regarding the minimum number of random start loading matrices that should be used for GPR-Varimax in PCA. We prefer to give conservative recommendations to ensure that a global optimum is detected for final solutions. It is possible that optimal solutions can already be found within fewer iterations of random start matrices for some conditions, but we use the number of iterations used in this study for recommendations. Also, the number of components that lie between the numbers investigated here (3, 6, 9, 12) are subsumed under the next larger number of components. Similarly, recommended sample sizes are regarded as minimum sample size to obtain the desired results, even though we did not perform our analyses for other sample sizes. For the rotation of up to three components in samples of $n = 100$ or larger, random start loading matrices should be iterated at least 10 times when GPR-Varimax is performed. For four to six components, we recommend using at least 50 iterations of random start matrices to find optimal solutions. If sample size is $n = 300$ or larger and loadings are Kaiser-normalized, 10 iterations are enough when four to six components are rotated. When more than six components are rotated, we recommend using Kaiser-normalized loadings because they reach optimal results within fewer iterations than non-normalized loadings. Moreover, sample size should be at least $n = 300$ for the rotation of more than six components. Kaiser-normalized random start loadings should be iterated at least 50 times in samples of $n \geq 300$ when seven, eight, or nine components are rotated. For 10 to 12 components, we recommend performing GPR-Varimax with at least 500 different random start loading matrices only in samples of $n = 300$ or larger and with Kaiser-normalized loadings to ensure optimal performance.

### 4.3   Future research

Whether 500 iterations suffice for the rotation of more than 12 components should be investigated in future research, as well as rotation performance in larger samples. Moreover, it would be interesting to extend the investigation from perfect simple structure to more complex loading structures in the population. For example, one could vary the size of the main loadings (here $a = .50$ for population factors/$a^* = .61$ for population components). Furthermore, one could introduce secondary loadings in addition to the main loadings. These could be modeled to maintain orthogonality (e.g., with alternating signs) or to produce obliqueness in the population model. This would be particularly interesting because in most cases, obliqueness is more likely than orthogonality despite the popularity of Varimax. For different conditions of obliqueness, GPR could be investigated for oblique rotation with Promax, Oblimin/Quartimin, or also Geomin as a newer method for oblique and orthogonal rotation (Browne, 2001). In addition to other rotation criteria, GPR could be applied to loadings from different extraction methods other than PCA to also examine GPR in factor rotation. For example, GPR could be applied to loadings obtained from maximum likelihood factor analysis or principal axis factoring. Furthermore, the present investigation focused on SPSS as one of the most popular commercial statistical software packages for factor analysis. We found that, when GPR-Varimax reached stationarity, results also reached the performance of the built-in procedure for Varimax rotation in SPSS. This finding provides additional support for the use of GPR in factor rotation, because it shows that the new algorithm for GPR can produce solutions like well-established software. One of the appeals of GPR is its implementation in the non-commercial software R (Bernaards & Jennrich, 2005), making it easily accessible for everyone who wants to use it. If comparisons like the present one with built-in procedures in commercial software show that GPR can compete with them, this might encourage the scientific community to switch to non-commercial software tools that are based on GPR. In line with this thought, of the abovementioned 41 hits on GoogleScholar for the combination of "gradient projection" and "Varimax", 25 of them comprised a combination of "gradient projection", "Varimax", and "R project" (status on September 13, 2018). This could indicate that GPR-Varimax is mainly used in R, whereas most scientists that use commercial software stick to pushing buttons to perform built-in, easily available procedures. Future investigations on GPR could include its implementations in R to extend the findings of the present study and provide additional support for its use in non-commercial software as well. On top of investigating the GPR algorithm as it exists at present, it is possible to adapt the algorithm itself by considering the findings of this study. Instead of iterating across a whole predefined set of numbers of iterations (here: 1, 10, 50, 100, 500, 1,000), one could adapt the algorithm to stop as soon as it reaches stationarity. The algorithm could begin with 10 iterations of start loading matrices, then use a next higher number (e.g. 50), compare the Varimax criterion between them, and either decide to increase the number of iterations or to stop if stationarity is reached. The criterion for stationarity of the Varimax criterion as we defined in the present study (differences of $\leq .0001$) could be used as a criterion to stop the search for better solutions. Congruences with population loadings are usually not available in empirical studies, but we found that they paralleled the results for stationarity of the Varimax criterion. To further improve the search for an optimal solution, one could switch to rational start loading matrices instead of random start loading matrices, which we used in this study. Rational start loadings are those obtained by a previous rotation. They are also recommended in an adaptation of GPR algorithms to the SCoTLASS problem, where factor extraction and Varimax rotation are joined in a simultaneous procedure (Trendafilov & Jolliffe, 2006).

### 4.4   Conclusion

To conclude, our study shows that GPR is a promising tool for component rotation. We identified conditions for optimal performance of GPR towards the Varimax criterion in PCA and give recommendations for scientists who wish to recover orthogonal simple structure with GPR-Varimax. To find optimal results for GPR-Varimax in PCA, we recommend using random start loadings instead of the original unrotated start loadings (i.e. where the start transformation matrix is identity). As a rule of thumb, random start loading matrices should be iterated at least 50 times for the rotation of up to six components in samples of $n = 100$ or larger. For the rotation of more than six components, sample size should be $n = 300$ or larger, loadings Kaiser-normalized, and random start loadings should be iterated at least 500 times to ensure optimal results. Future simulations, which include more complex loading structures and other rotation and extraction methods, could further support the usage of GPR and make it a more common tool for factor and component rotation in the scientific community.

**Supplement: Results for the root mean square error (RMSE)**

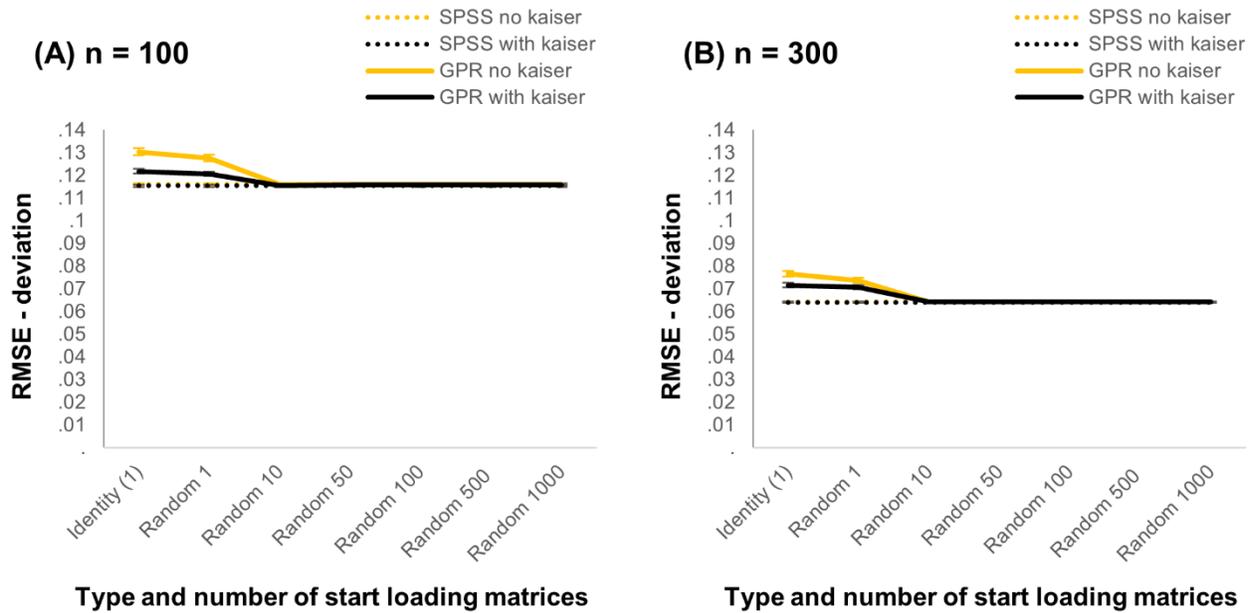

*Supplementary Figure 1.* Mean (and standard error) of the root mean square error (RMSE) as deviation of rotated components from population components for the rotation of 3 components. *Type of start loadings* refers to the start transformation matrix **T** by which the unrotated loadings **A** are post-multiplied before GPR-Varimax rotation. *Identity* means that **A** was inserted without alterations. The maximum value of the axis is the maximum empirical value for mean RMSE from the simulation + .01.

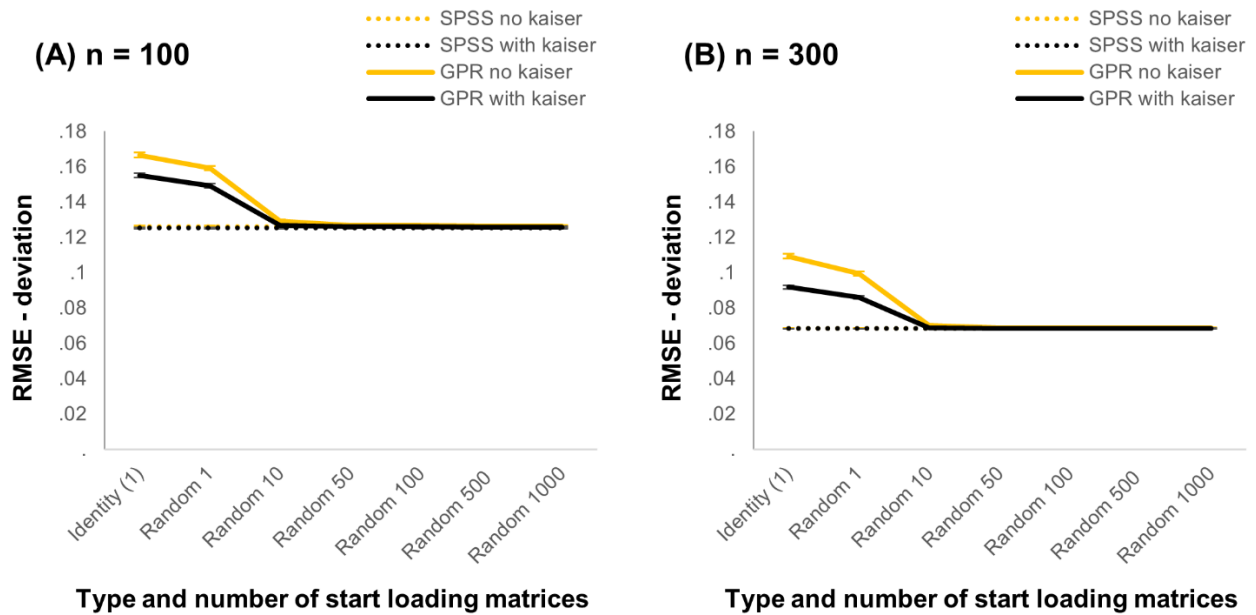

*Supplementary Figure 2.* Mean (and standard error) of the root mean square error (RMSE) as deviation of rotated components from population components for the rotation of 6 components. *Type of start loadings* refers to the start transformation matrix **T** by which the unrotated loadings **A** are post-multiplied before GPR-Varimax rotation. *Identity* means that **A** was inserted without alterations. The maximum value of the axis is the maximum empirical value for mean RMSE from the simulation + .01.

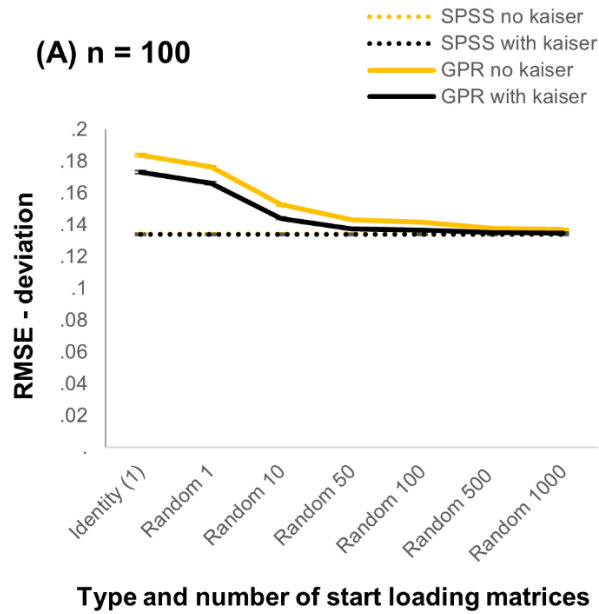
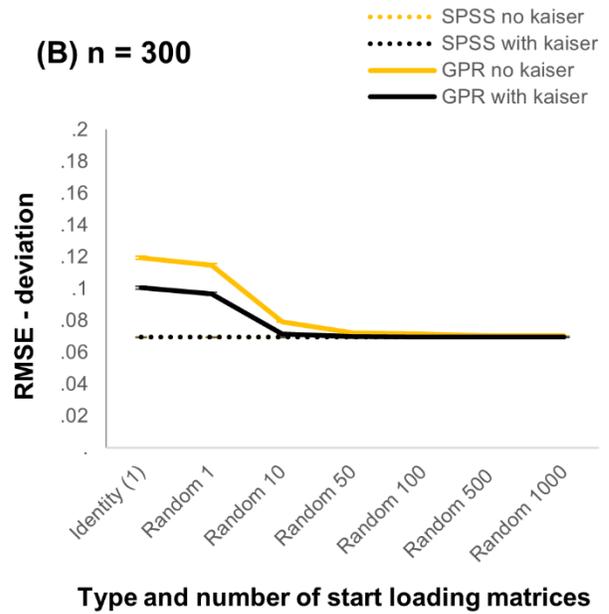

*Supplementary Figure 3.* Mean (and standard error) of the root mean square error (RMSE) as deviation of rotated components from population components for the rotation of 9 components. *Type of start loadings* refers to the start transformation matrix **T** by which the unrotated loadings **A** are post-multiplied before GPR-Varimax rotation. *Identity* means that **A** was inserted without alterations. The maximum value of the axis is the maximum empirical value for mean RMSE from the simulation + .01.

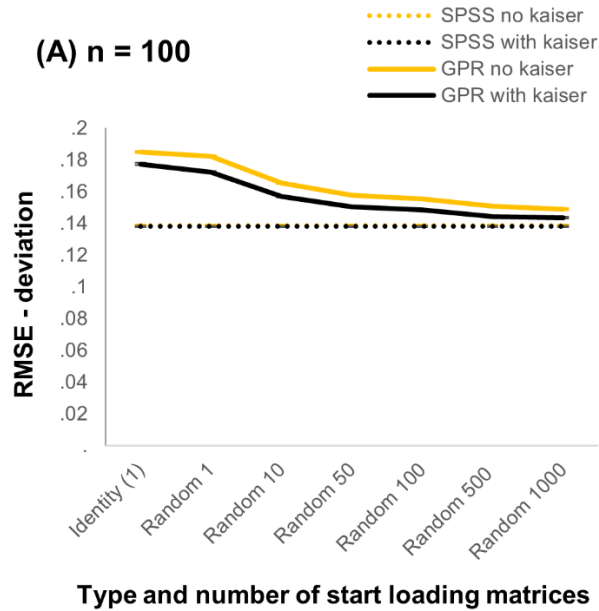
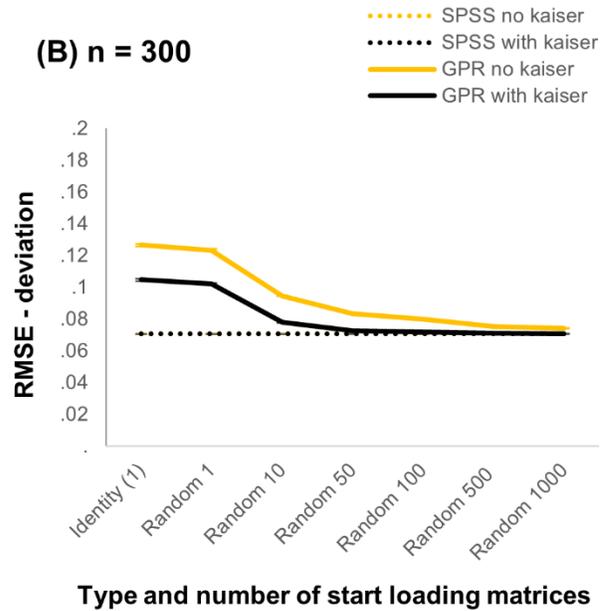

*Supplementary Figure 4.* Mean (and standard error) of the root mean square error (RMSE) as deviation of rotated components from population components for the rotation of 12 components. *Type of start loadings* refers to the start transformation matrix **T** by which the unrotated loadings **A** are post-multiplied before GPR-Varimax rotation. *Identity* means that **A** was inserted without alterations. The maximum value of the axis is the maximum empirical value for mean RMSE from the simulation + .01.